\documentclass[aps,prl,showpacs,twocolumn,superscriptaddress]{revtex4}
\usepackage{xy,amsmath,amssymb,amsthm,graphics,graphicx,bm,subfigure,epstopdf,units,natbib}
\usepackage[colorlinks,linkcolor=red]{hyperref}
\newcommand{\arxiv}[2][arxiv:]{\href{http://arxiv.org/abs/#1#2}{#1#2}}

\newcommand{\be}{\begin{equation}}
\newcommand{\ee}{\end{equation}}
\newcommand{\ben}{\begin{eqnarray}}
\newcommand{\een}{\end{eqnarray}}
\newcommand{\bes}{\begin{subequations}}
\newcommand{\ees}{\end{subequations}}
\newcommand{\bF}{\begin{figure}}
\newcommand{\eF}{\end{figure}}

\def\ket#1{ | #1 \rangle}

\begin{document}

\title{Multi-photon state engineering by heralded interference between single photons and coherent states}

\author{Tim J. Bartley}
\email{t.bartley1@physics.ox.ac.uk}
\affiliation{Clarendon Laboratory, Department of Physics, University of Oxford, Oxford OX1 3PU, United Kingdom}

\author{Gaia Donati}
\affiliation{Clarendon Laboratory, Department of Physics, University of Oxford, Oxford OX1 3PU, United Kingdom}

\author{Justin B. Spring}
\affiliation{Clarendon Laboratory, Department of Physics, University of Oxford, Oxford OX1 3PU, United Kingdom}

\author{Xian-Min Jin}
\affiliation{Clarendon Laboratory, Department of Physics, University of Oxford, Oxford OX1 3PU, United Kingdom}
\affiliation{Centre for Quantum Technologies, National University of Singapore, 117543, Singapore
}

\author{Marco Barbieri}
\affiliation{Clarendon Laboratory, Department of Physics, University of Oxford, Oxford OX1 3PU, United Kingdom}


\author{Animesh Datta}
\affiliation{Clarendon Laboratory, Department of Physics, University of Oxford, Oxford OX1 3PU, United Kingdom}


\author{Brian J. Smith}
\affiliation{Clarendon Laboratory, Department of Physics, University of Oxford, Oxford OX1 3PU, United Kingdom}

\author{Ian A. Walmsley}
\affiliation{Clarendon Laboratory, Department of Physics, University of Oxford, Oxford OX1 3PU, United Kingdom}

\date{\today}

\begin{abstract}

We develop a technique for generating multi-photon nonclassical states via interference between coherent and Fock states using quantum catalysis. By modulating the coherent field strength, the number of catalyst photons and the ratio of the beam splitter upon which they interfere, a wide range of nonclassical phenomena can be created, including squeezing of up to \unit[1.25]{dB}, anti- and super-bunched photon statistics and states exhibiting over 90\% fidelity to displaced coherent superposition states. We perform quantum catalysis experimentally, showing tunability into the nonclassical regime. Our protocol is not limited by weak nonlinearities that underlie most known strategies of preparing multi-photon nonclassical states. Successive iterations of this protocol can lead to direct control over the weights of higher-order terms in the Fock basis, paving the way towards conditional preparation of ``designer'' multi-photon states for applications in quantum computation, communication and metrology.

\end{abstract}



\maketitle

\section{Introduction}\label{sec:Intro}

Quantum technologies promise enhanced performance of computation, communication, sensing, and simulation protocols. Photonic quantum information processing has been immensely successful in harnessing these advantages in several settings~\cite{Kok07,OBr09,Kok10}. This is because the quantum states of photons can be prepared, processed and measured with ease and precision whilst simultaneously preserving fragile quantum phenomena, even in a hostile environment as they do not suffer from any significant coupling with the external environment. Depending on the optical degrees of freedom in which quantum information is stored, manipulated and detected, photonic quantum information processing typically operates in two different regimes: discrete variable (DV) and continuous variable~\cite{Bra05,And10,Wee11} (CV). In contrast to DV states residing in finite-dimensional Hilbert spaces, the CV quadratures of an optical field allow the encoding of information over infinite dimensions. This approach has already demonstrated advantages in the implementation of secure key distribution~\cite{Gro03}, and there are ongoing efforts to exploit its potential in computation and communication~\cite{Lun08,Nee10,Dat12}. For a protocol to provide quantum advantages not attainable classically, it is crucial that the quantum states involved exhibit non-Gaussian features in their Wigner distributions in phase space~\cite{Llo99,Bar02,Nis09}.

Generating non-Gaussian states normally requires nonlinearities of the third order in the field operators~\cite{Bra05}. However, such effects are negligible at low photon flux. In this regime---the most relevant scenario in current experiments at the quantum level---a more feasible strategy involves the use of conditional probabilistic operations~\cite{Lvo02}. These comprise the manipulation of the quantum state by means of linear optical elements~\cite{Sch05}, which generates an effective nonlinearity by accepting only particular outcomes of measurements on ancillary modes. Such schemes have allowed the production of superpositions of coherent states~\cite{Our07, Bla12}, and manipulation at the single-photon level~\cite{San06,Res07,Par07,Fer10,Xia10}.
\begin{figure}[b]
\includegraphics[scale=1]{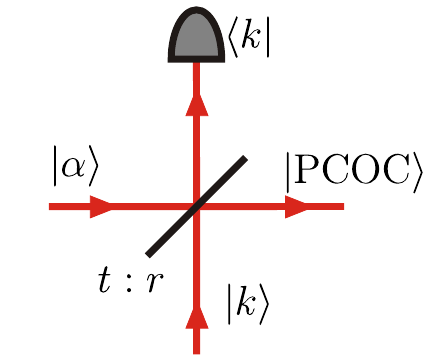}
\caption{A Photon Catalysed Optical Coherent state $\ket{\textrm{PCOC}}$ is produced following the interference  between a coherent state $\ket{\alpha}$ and a $k$-photon Fock state $\ket{k}$ (at a beam splitter of reflectivity $|r|^2=1-|t|^2$), conditional on measuring $k$ photons at one output port.}\label{fig:PCOC}
\end{figure}

\begin{figure*}[!bthp]
\centering
\subfigure[]{\includegraphics[scale=0.7]{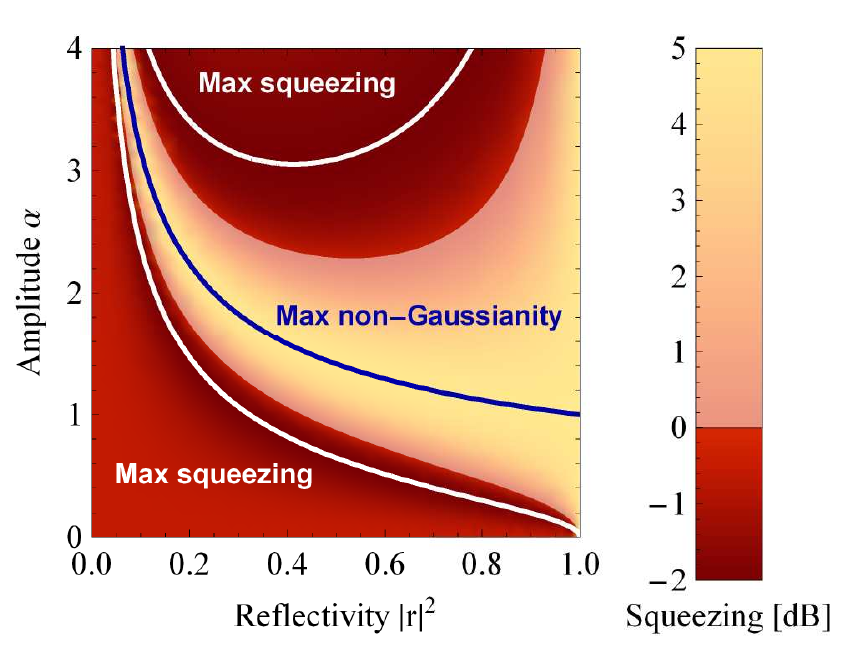}\label{fig:SqReg}}\qquad
\subfigure[]{\includegraphics[scale=0.7]{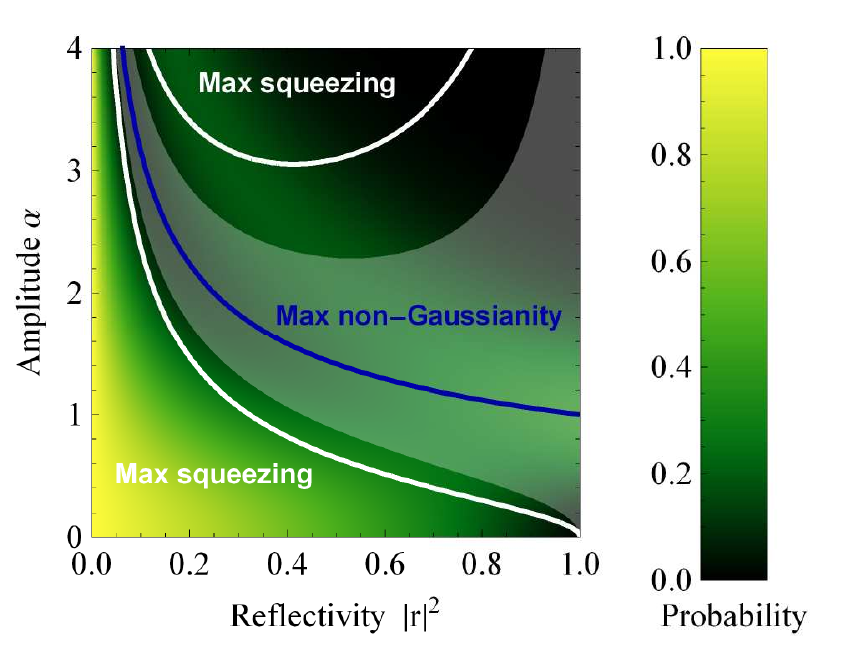}\label{fig:SqProb}}\qquad
\subfigure[]{\includegraphics[scale=0.7]{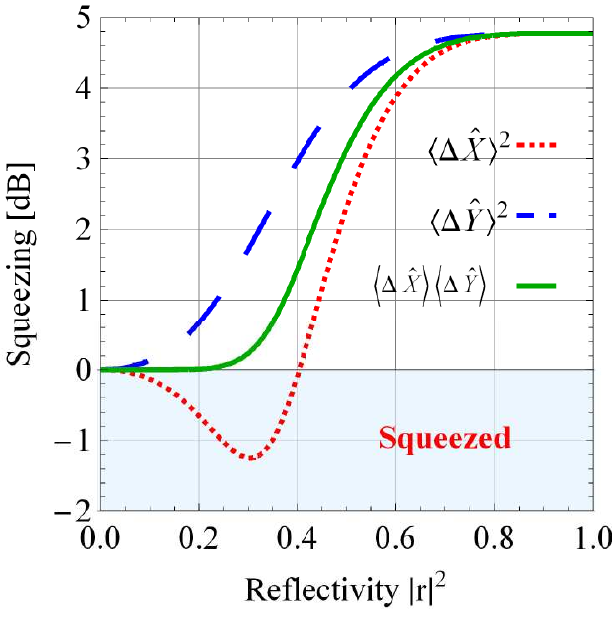}\label{fig:SqAlp1}}
\caption{Squeezing behaviour as a function of interaction parameters $|r|^2$ and $\alpha$. The range of parameter space that generates squeezing is shown in \subref{fig:SqReg}, with the probability of successful catalysis shown in~\subref{fig:SqProb}. The white lines show the parameters which generate the maximum squeezing of \unit[1.25]{dB} below the shot-noise limit and the blue lines show the maximum non-Gaussianity, corresponding to a deviation from a minumum uncertaity state of \unit[4.77]{dB}. \subref{fig:SqAlp1} shows the quadrature variances, relative to the vacuum (unsqueezed) state in units of dB, at a coherent state amplitude of $\alpha=1$.  Red dotted line, blue dashed line: $\hat{X}$ and $\hat{Y}$ quadrature variance, respectively. Green solid line: joint variance.}
\end{figure*}
In this Article, we investigate the possibilities offered to multi-photon quantum state engineering by a particular interferometric scheme called quantum catalysis~\cite{Lvo02}, which uses a photonic Fock state to modulate the probability amplitudes of a coherent state.
The analogy to catalysis is motivated by conditioning on the same number of photons, say $k,$ as is initially interfered with the coherent state, as shown in Fig.~\ref{fig:PCOC}. We generalise the result in Ref.~\cite{Lvo02} (in which $k=1,~\alpha\ll1$ and beam splitter transmissivity $t\ll1$) across a wide range of parameters and to higher-order in the photon number distribution, introducing a broad family of multi-photon quantum states. We experimentally measure the second-order autocorrelation function $g^{(2)}\left(0\right)$ for states catalysed by one photon ($k=1$), as well as their photon-number distributions. The states we produce are deep in the quantum regime, and our strategy allows for the production of other nonclassical states such as squeezed and Schr\"odinger kitten states. In spite of injecting and extracting the same number of photons, the photon-number distribution of the final quantum state is radically modified due to quantum interference. This operates at two levels: firstly, each Fock state contributing to the coherent input state interferes with the catalysing $k$ photons, undergoing a generalised Hong-Ou-Mandel type interference~\cite{San06}. Secondly, the outcomes of these individual interferences then add coherently due to the inherent phase reference in a coherent state. This combination results in a completely novel class of multi-photon quantum states with a whole gamut of uniquely nonclassical properties.

\section{Introducing PCOC states}

%

The schematic for the generation of a photon-catalysed optical coherent (PCOC) state is shown in Fig.~(\ref{fig:PCOC}). The conditioned state at the output is given by
\begin{equation}\label{eqn:state}
\ket{\textrm{PCOC}}=\mathcal{N}_{\alpha,r}\sum_{n=0}^\infty\frac{\alpha^n}{\sqrt{n!}}C_{n}\left(r,t,k\right)\ket{n}~,
\end{equation}
where
\begin{equation}
C_{n}\left(r,t,k\right)=\sum_{j=0}^{\min\left(n,k\right)}\left(\begin{array}{c}
n\\n-j
\end{array}\right)\left(\begin{array}{c}
k\\j
\end{array}\right)\left(-1\right)^jt^{n+k-2j}r^{2j}~,
\end{equation}
$r,t$ are the reflectivity and transmissivity of the beam splitter, respectively (satisfying $|r|^2+|t|^2=1$) and $\mathcal{N}_{\alpha,r}$ is the normalisation constant. The states cover a range going from a coherent state ($r=0$) to a $k$-photon Fock state ($r=1$). 
By tuning the parameters of the interaction, namely the beam splitter reflectivity $r$, coherent state amplitude $\alpha$ and the number of catalyst photons $k$, the coefficients of the photon-number terms (Fock layers) may be modulated, generating a wide range of nonclassical phenomena, as we show next.

\begin{figure*}[!tbhp]
\centering
\subfigure[]{\includegraphics[angle=270]{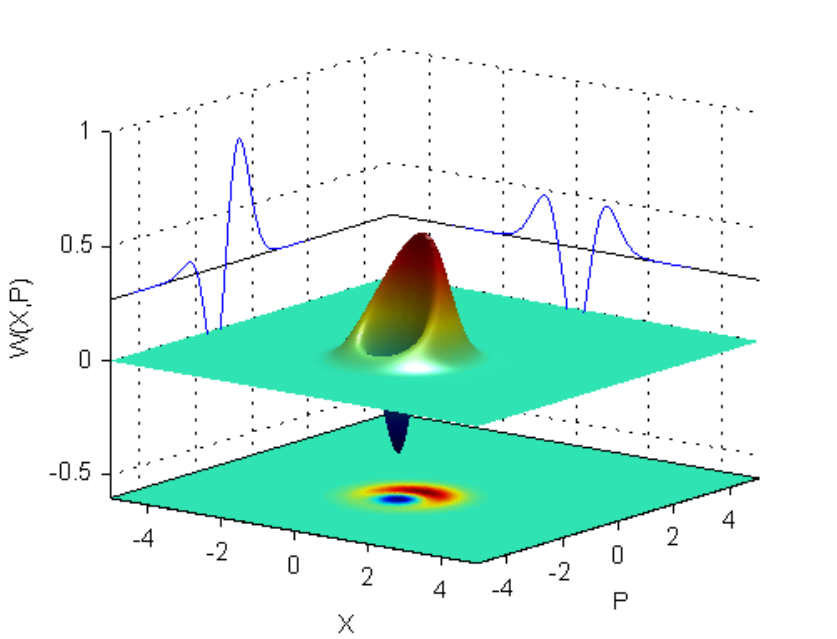}\label{fig:PCOCWig}}
\subfigure[]{\includegraphics[angle=270]{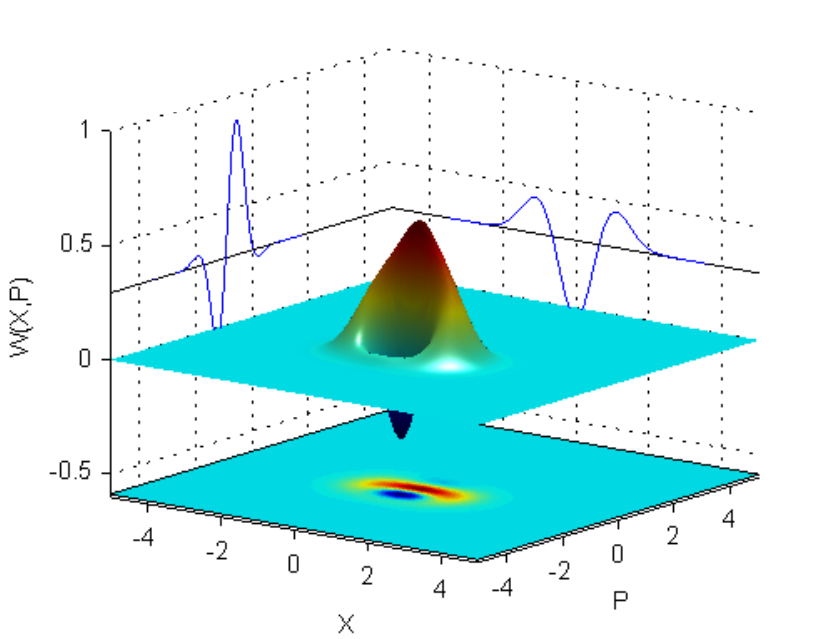}\label{fig:KittenWig}}
\caption{Wigner functions of the photon catalysed state \subref{fig:PCOCWig} and displaced coherent state superposition  (CSS)~\subref{fig:KittenWig}, of amplitude $\alpha=0.9$, displaced by $\beta=0.8$ from Eq.~(\ref{eqn:CSS}). Their Wigner functions are similar, indeed the photon-catalysed state exhibits a fidelity of 0.9 with respect to the displaced coherent state superposition. This result shows that CSS states can be generated without squeezing using the catalysis scheme.}
\label{fig:Wigners}
\end{figure*}

\begin{figure*}[!tbhp]
\centering
\subfigure[~$k=2$, $|r|^2=1/2$, $\alpha=2$]{\includegraphics[angle=270,scale=0.65]{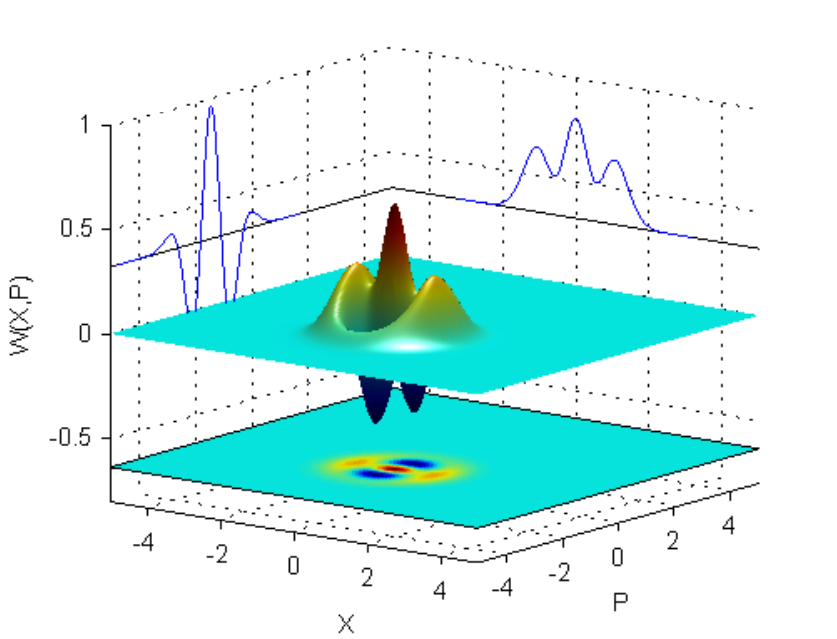}\label{fig:22Wig}}
\subfigure[~$k=3$, $|r|^2=1/4$, $\alpha=2$]{\includegraphics[angle=270,scale=0.65]{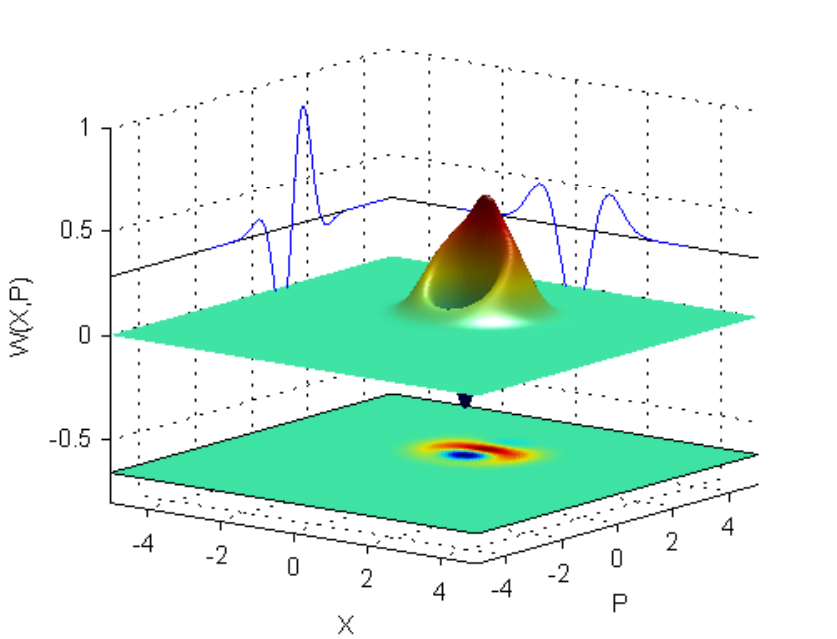}\label{fig:33Wig}}
\subfigure[~$k=6$, $|r|^2=4/9$, $\alpha=2.7$]{\includegraphics[angle=270,scale=0.65]{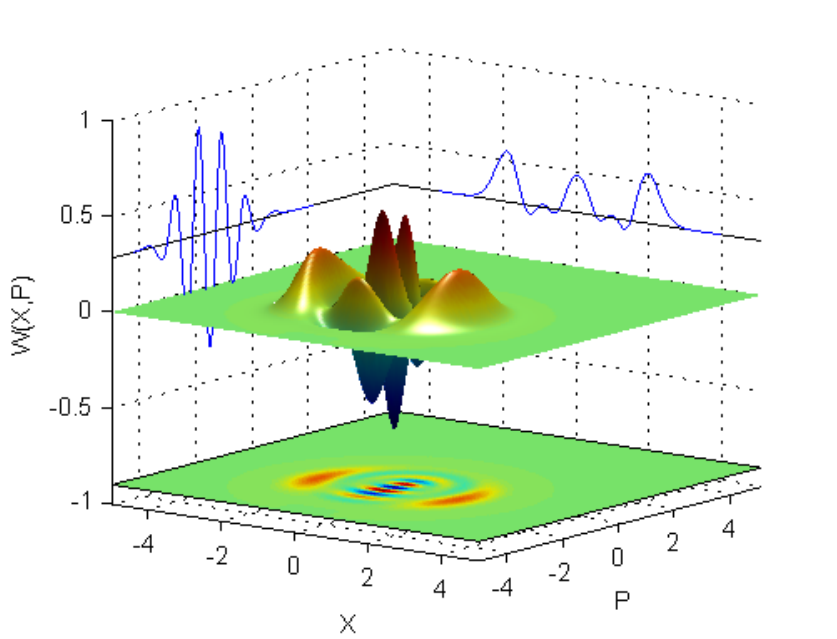}\label{fig:66Wig}}
\caption{Wigner functions of higher-order catalysed states showing the rich structure that can be achieved.}\label{fig:MoreWigs}
\end{figure*}

Squeezed states form an essential part of the quantum toolbox, and PCOC states can, for a range of parameters, yield an $\hat{X}$ quadrature variance $\Delta^2 \hat{X}$ below the standard limit. This is shown in the dark regions of Fig.~\ref{fig:SqReg}. Analytical expressions are obtained for the variance of the quadratures: 
\begin{widetext}
\begin{equation}
{\Delta^2 \hat{X}}=\frac{\left(1-r^2\right)^2-4 r^2 \left(1-r^2\right)^2 |\alpha| ^2+3 r^4 \left(2-4 r^2+3 r^4\right) |\alpha| ^4-4 r^6 \left(1-r^2\right)^2 |\alpha| ^6+r^8 \left(1-r^2\right)^2 |\alpha| ^8}{4 \left(1-r^2 \left(1+|\alpha| ^2 \left(2-r^2 \left(3+\left(1-r^2\right) |\alpha| ^2\right)\right)\right)\right)^2}~,
\end{equation}
and
\begin{equation}\label{eq:Xvar}
\Delta^2 \hat{P}=\frac{1}{4}+\frac{r^4 |\alpha| ^2}{2 \left(1-r^2 \left(1+|\alpha| ^2 \left(2-r^2 \left(3+\left(1-r^2\right) |\alpha| ^2\right)\right)\right)\right)}~.
\end{equation}
\end{widetext}
These quadrature variances, relative to their vacuum values of $1/4$, are plotted on a logarithmic scale (units of dB) in Figs.~2(a)  and~2(c) (for $\alpha=1$) showing the range of squeezing for $\alpha$ and $|r|^2$.

By minimizing the expressions above for $\Delta^2 \hat{X}$, the largest squeezing is attainable is $3/16$, below the vacuum noise level of $1/4$ by \unit[1.25]{dB}. The maximum variance of $3/4$ corresponds to \unit[4.77]{dB} antisqueezing. The parameters that yield these values are found by solving $\Delta^2 \hat{X}{=}3/16$ under the constraints $0\leq r\leq1,|\alpha|<0$ (using Mathematica). The solutions to this equation are not uniquely valued: they form two continuous lines in $\left(|r|^2,\alpha\right)$ parameter space, as shown in Fig.~2(a). The equations corresponding to the loci of minima are
\begin{equation}
\alpha_\textrm{min}=\sqrt{\frac{\sqrt{3\left(4-r\right)r}\pm \left(2+r\right)}{2r\left(1\mp r\right)}}~.
\end{equation}
Similarly for the locus of maximum non-Gaussianity is described by
\begin{equation}
\alpha_\textrm{max}=\frac{1}{\sqrt{r}}~.
\end{equation}
Reaching the maximum squeezing and non-Gaussianity by this scheme is therefore independent of the initial coherent state amplitude $\alpha$.

However, the amount of squeezing obtainable, particularly at high $\alpha$, is offset by the  probability of successful catalysis: if the coherent state is large, the probability of detecting one photon is small. The probability of successful catalysis is given by
\begin{equation}
\begin{split}
&P_\textrm{PCOC}\left(\alpha,r\right)=\sum_{n=0}^\infty|c_{n}\left(\alpha,r\right)|^2\\
&{=}e^{-r^2 |\alpha| ^2} \left(1-r^2 \left(1-|\alpha| ^2 \left(r^2 \left(3+|\alpha| ^2\right)-2-r^4 |\alpha| ^2\right)\right)\right)
\end{split}
\end{equation}
and it is shown in Fig.~\ref{fig:SqProb}. For low $\alpha$, catalysis is fairly likely. Indeed, at $\alpha=1$ and $|r|^2=0.332$, which generates maximum squeezing, the probability of successful catalysis is about 47\%.

The behavior of both quadrature variances at $\alpha=1$ is shown in Fig.~\ref{fig:SqAlp1}. The $\hat{P}$ quadrature is never squeezed, and the combined variances show that the state clearly satisfies the minimum variance condition. For $|r|^2 \lesssim 0.2$, the state asymptotically approaches a minimum uncertainty state, whereas this is lost for $|r|^2 \gtrsim 0.2$. Since the state is pure, this means states in this region are non-Gaussian, implying negativity in their Wigner distributions~\cite{Hud74}.

Non-Gaussian multi-photon states have been shown to be useful for  computation~\cite{Nis09,Jeo05} and communication~\cite{Eis02}. One class of non-Gaussian states of particular interest are coherent state superposition  (CSS) states. As a superposition of coherent state amplitudes separated in phase space, they are the optical analog of the Schr\"odinger cat ``dead'' and ``alive'' state~\cite{Sch35}. When the separation between the amplitudes is small, the states are termed Schr\"odinger kitten states~\cite{Our06}, and it has been shown that combining small CSS states can yield much larger CSS states necessary for universal quantum computation~\cite{Lun04}.

A CSS state of amplitude $\alpha$, displaced in phase space by $\hat{D}\left(\beta\right)$, is given by
\begin{equation}\label{eqn:CSS}
\ket{\textrm{CSS}}=\mathcal{N}\sum_{n=0}^\infty\frac{\left(\beta+\alpha\right)^n+\left(\beta-\alpha\right)^n}{\sqrt{n!}}~\ket{n}.
\end{equation}
An example of a weak CSS state ($\alpha=0.9$), displaced in phase space ($\beta=0.8$), is shown in~\ref{fig:KittenWig}.
With a particular choice of interaction parameters ($|r|^2=0.77,~\alpha=1.35$), the PCOC state approximates a displaced weak coherent superposition state,
with $>90\%$ fidelity, with a Wigner function as shown in Fig.~\ref{fig:PCOCWig}, which is clearly non-Gaussian. Previous schemes have typically required some initial squeezing~\cite{Dak97,
Our06,Nee06,
Tak08,Ger10}; our results indicate a route to these states without such requirements.

Simple extensions of the catalysis scheme allow for the preparation of more sophisticated quantum states. A catalyst of $k\geq 2$ photons in Eq.~(\ref{eqn:state}) leads to more elaborate Wigner functions  as shown in Fig.~(\ref{fig:MoreWigs}). A further extension is to concatenate iterations of catalysis. Each iteration $i$ introduces an additional reflectivity $r_i$, such that the total reflectivity $\mathbf{r}=\left(r_1,\ldots,r_l\right)$. The state following $l$ iterations may be written
\begin{equation}
\ket{\textrm{PCOC}}_l=\mathcal{N}_{\alpha,\mathbf{r}}\sum_{n=0}^\infty\frac{\alpha^n}{\sqrt{n!}}\prod_{i=0}^{l}c_{n,i}\ket{n}~,
\end{equation}
where
\begin{equation}
c_{n,i}=\sum_{j=0}^{\min\left(n,k_i\right)}
\left(\begin{array}{c}
n\\n-j
\end{array}\right)\left(\begin{array}{c}
k_i\\j
\end{array}\right)
\left(-1\right)^jt_i^{n+k_i-2j}r_i^{2j}~.
\end{equation}
Such an iterative scheme forms the basis of a number of conditional approaches towards quantum computation~\cite{Sch05}. The coefficients of the Fock layers can be directly controlled by the reflectivities at each iteration~\cite{TB}. As such, arbitrary quantum states can be generated by heralding the successful catalysis interaction on a single mode, without invoking entanglement~\cite{Bim10}.

\begin{figure}[bth]
\centering
\includegraphics[width=0.48\textwidth]{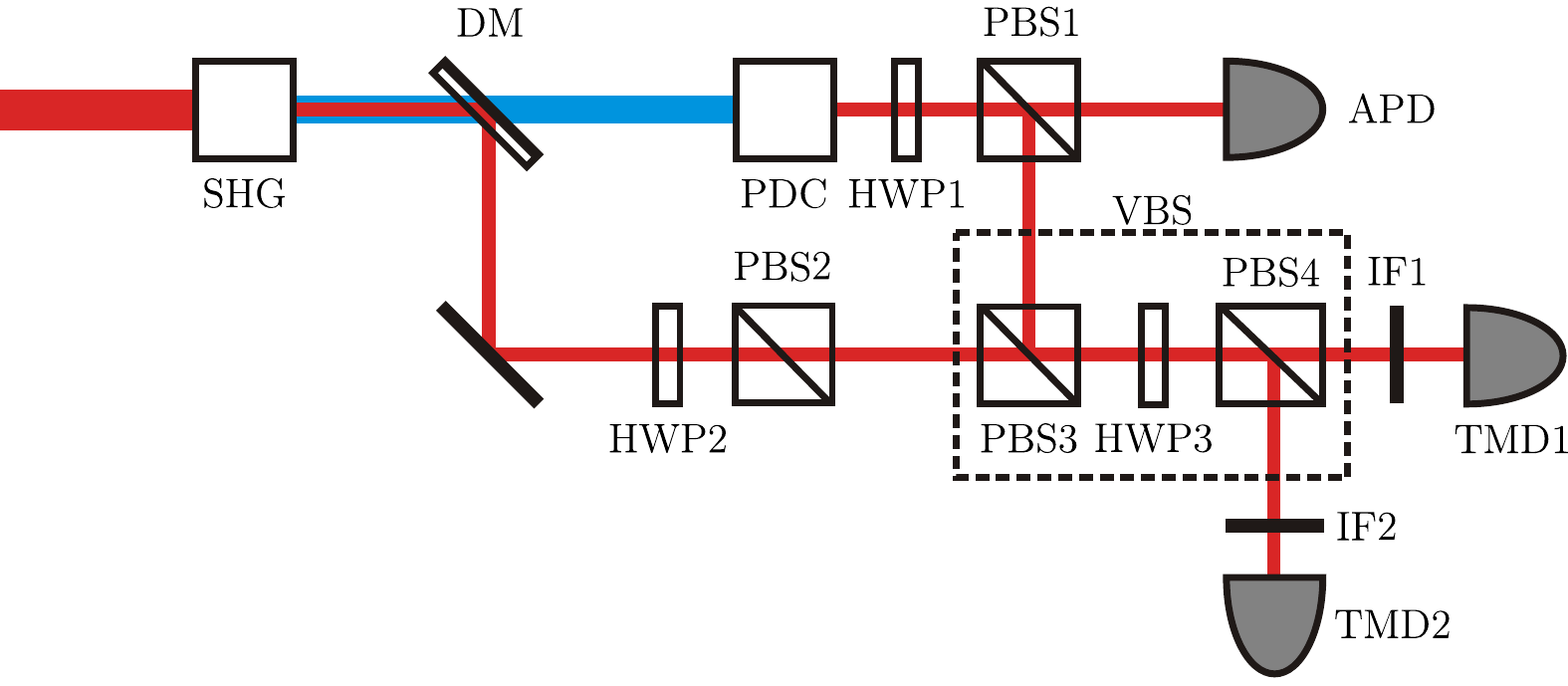}
\caption{Experimental setup for probing single-photon catalysis. A Ti:Sapphire laser is doubled by a second harmonic generation crystal (SHG). The up-converted beam undergoes type-II co-linear parametric down-conversion (PDC), producing pairs of orthogonally polarized photons in pure spectral-temporal states~\cite{Mos08}. HWP1 ensures that the $H$-polarized photon is detected by an avalanche photodiode (APD), which heralds the presence of a photon in the $V$ polarization mode.  The residual fundamental, which is used as the coherent state, is reflected by the dichroic mirror DM. This beam is then attenuated by a half-wave plate HWP2 and polarizing beam splitter PBS2 to single-photon level intensity. The two beams are spatially recombined on PBS3, and synchronized by adjusting the position of a delay stage. This operation is followed by interference on the variable beam splitter constituted by HWP3 and PBS4. The two polarization modes are then filtered by \unit[3]{nm} interference filters (IF), coupled to single-mode fibres and sent to photon-number-resolving time-multiplexed detectors (TMDs) for photon counting~\cite{Ach04}.}\label{fig:setup}
\end{figure}
\section{Experimental tuning of the second-order coherence}
To demonstrate the practical efficacy of our proposal, we experimentally generate PCOC states catalysed by one photon. Using the setup shown in Fig.~(\ref{fig:setup}), we measure the photon number distribution of the PCOC state at a range of beam splitter reflectivities. The variable beam splitter comprises a PBS (PBS3 - to spatially overlap the coherent state and Fock states), a half-wave plate (HWP3) and interference beam splitter (PBS4). The beam splitter reflectivity and half-wave plate angle in the variable beam splitter are related by  $r=1+\cos(2 \theta)$. 

\begin{figure}[t]
\subfigure[~Experimental data]{\includegraphics[angle=0,width=0.48\textwidth]{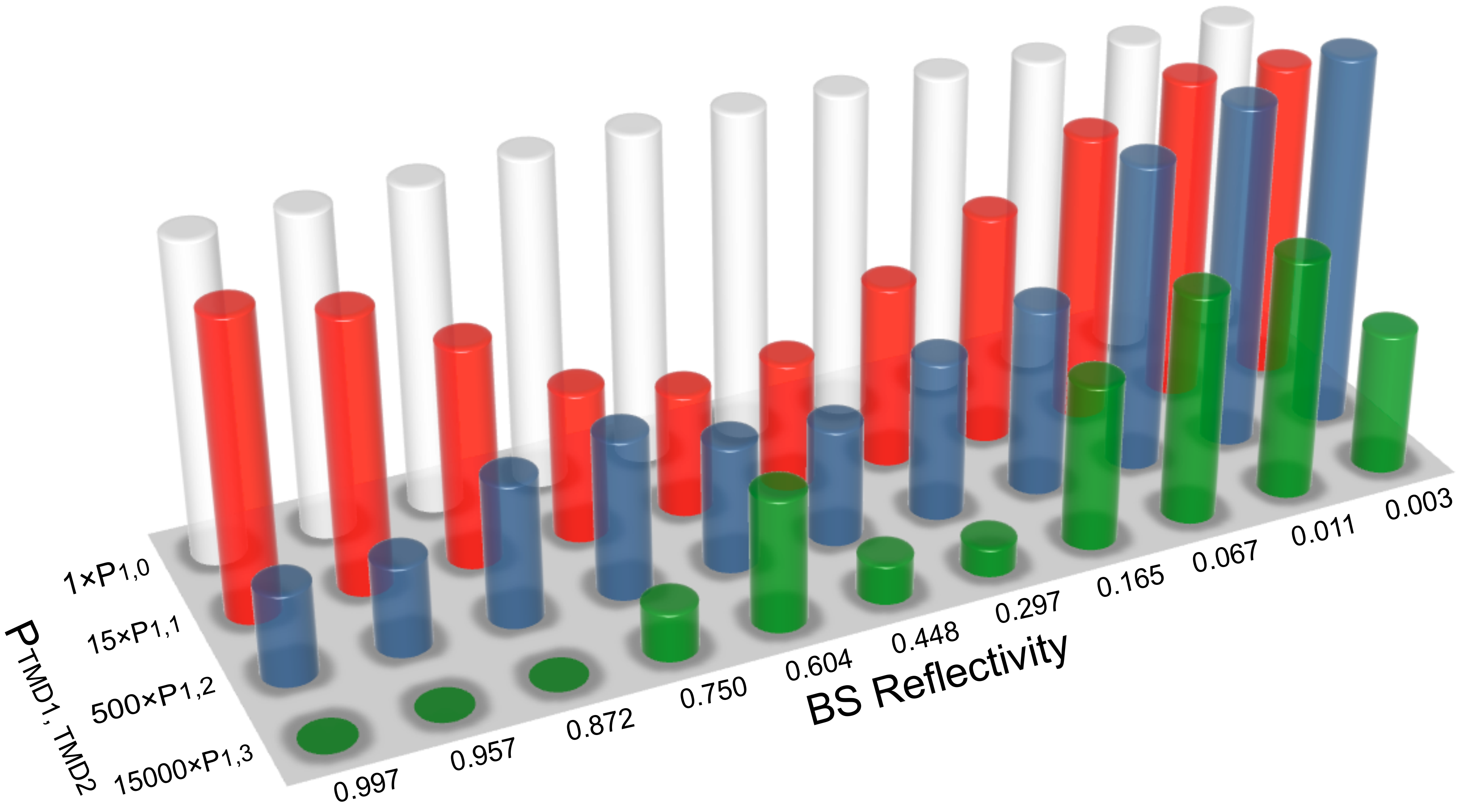}\label{fig2:exp}}
\subfigure[~Theoretical reconstruction]{\includegraphics[angle=0,width=0.48\textwidth]{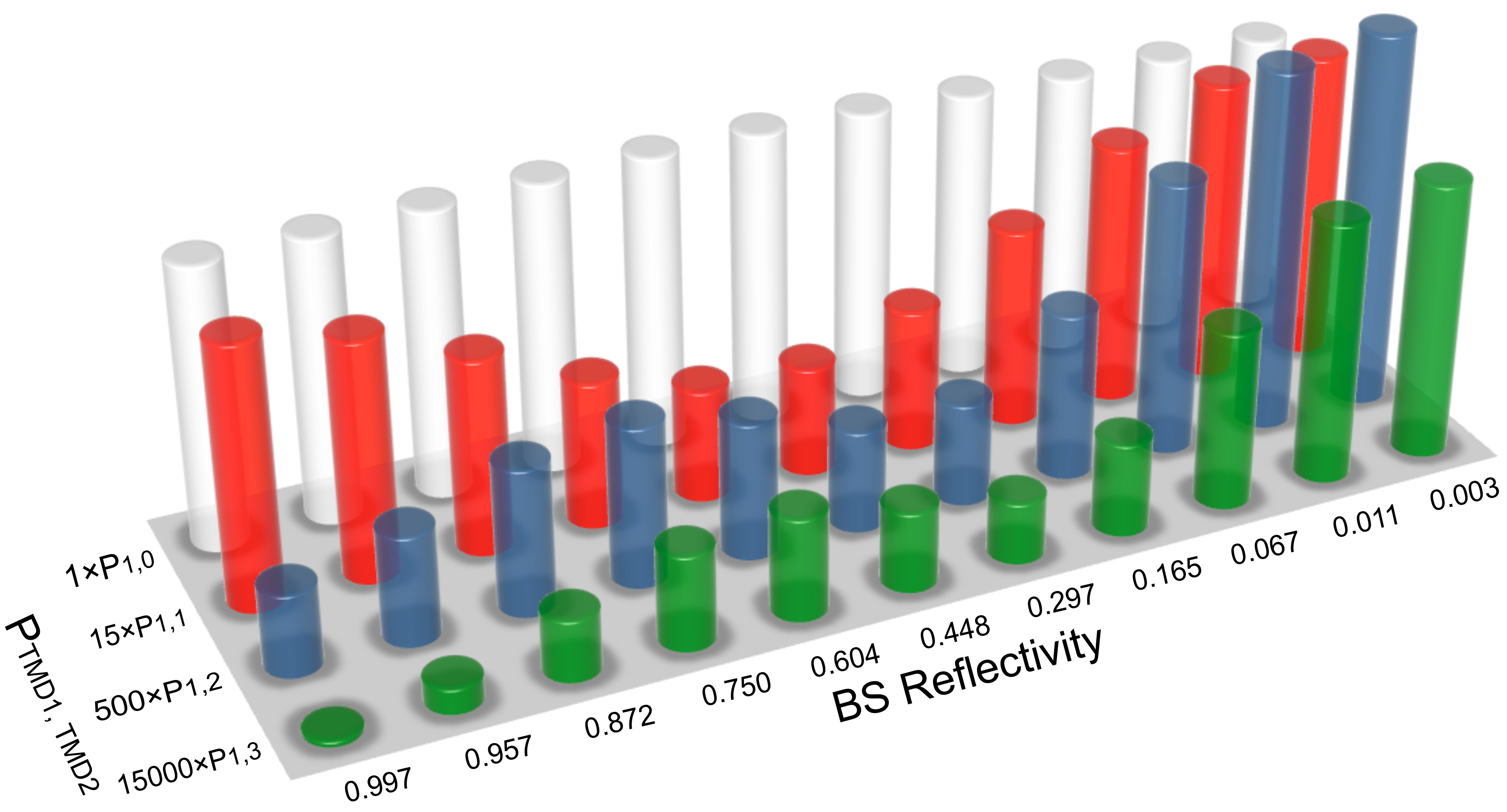}\label{fig2:theory}}
\caption{Measured~\subref{fig2:exp} and theoretical~\subref{fig2:theory} photon number distributions as a function of beam splitter reflectivity $|r|^2$ for a coherent state mean photon number of $\left|\alpha\right|^2=1.11$. White shows the vacuum component, while red, blue and green represent the one-, two- and three-photon terms, respectively. 
Note that the higher-order terms have been scaled for clarity; the captions $X\times P_{i,j}$ corresponds to a scaling factor $X$ multiplied by the probability $P$ of a click corresponding to $i$ photons detected in TMD1 and $j$ photons detected in TMD2.
}\label{fig:dists}
\end{figure}

The resulting modes after interference are sent to the detector array. By multiplexing APDs spatially and temporally, we are able to achieve photon number resolution up to eight photons plus vacuum~\cite{Ach04}. Typical results from number-resolved counting are shown in Fig.~(\ref{fig:dists}), taking into account the binning of our detector. The action of catalysis as a Fock state filter appears in the modulation of the higher-order photon terms as the reflectivity is tuned~\cite{San06}. The single photon component is suppressed with respect to what one would expect in a coherent state. A similar effect is also present in the two- and three-photon components, albeit with a lower contrast.

\begin{figure}[thb]
\includegraphics[angle=0,width=0.48\textwidth]{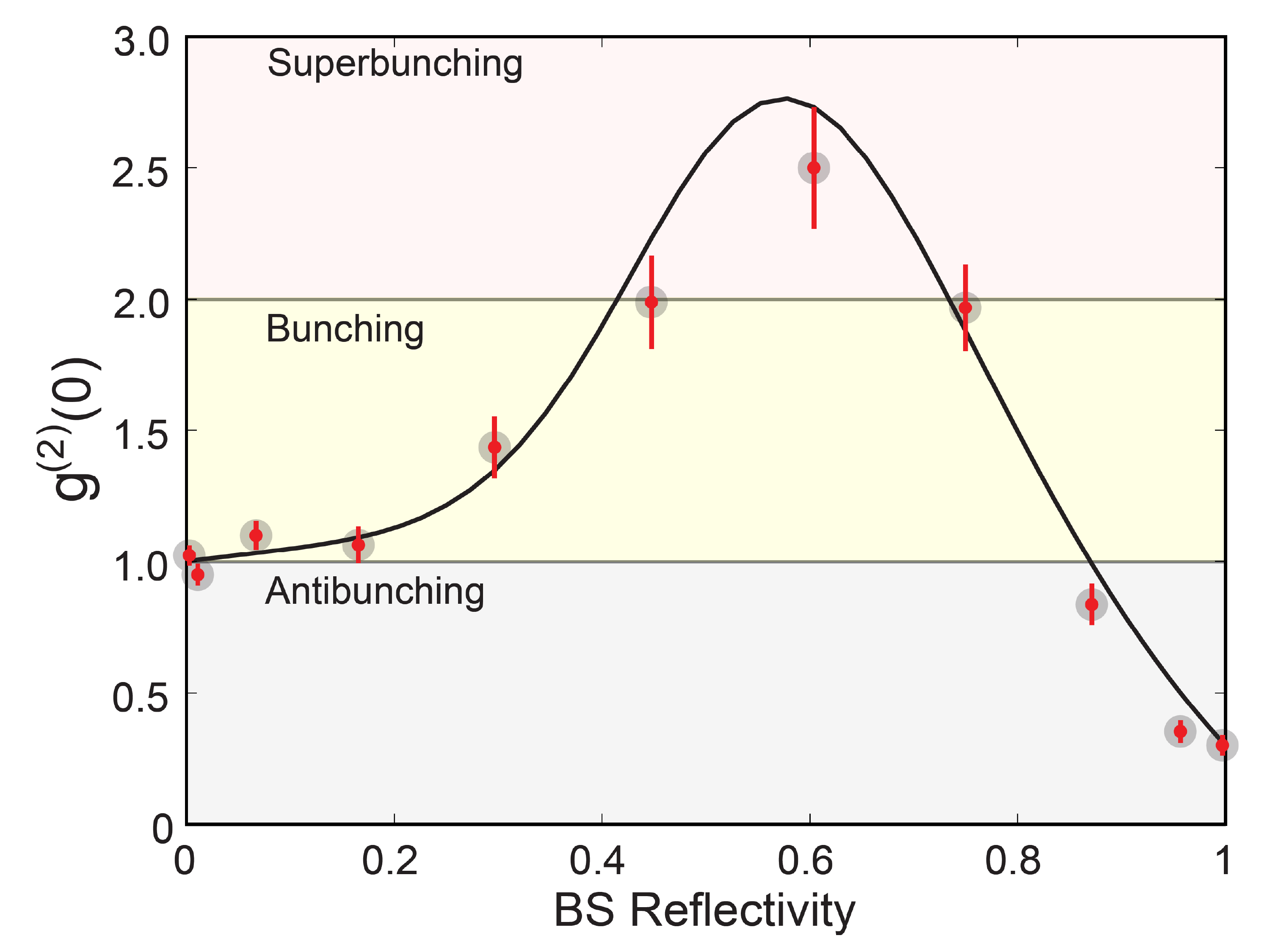}
\caption{Second order autocorrelation function $g^{(2)}(0)$ as a function of beam splitter reflectivity, for a coherent state mean photon number of $\left|\alpha\right|^2=1.11$. The points are measured values based on the photon-statistics in Fig.~\ref{fig2:exp}, while the line is the expected value, incorporating inefficiency in our detection system. For $|r|^2=0,1$, $g^{(2)}(0)$ approaches the values for a coherent state and single photon, respectively. The regions showing anti-bunching $g^{(2)}(0)<1$ (strictly nonclassical), bunching $1\leq g^{(2)}(0)\leq2$ and super-bunching $g^{(2)}(0)>2$ are clear.}
\label{fig:g2}
\end{figure}

We also measure the second order autocorrelation function $g^{\left(2\right)}\left(0\right)$ and show its variation with beam splitter reflectivity $|r|^2$.  Fig.~(\ref{fig:g2}) shows that states lying below the classical bound ($g^{(2)}(0)=1$) are observed for high reflectivities. In this region, our experiment realizes a displacement of a single photon, as first observed in~\cite{Lvo02}. This signature of nonclassicality disappears rather quickly as $|r|^2$ decreases below 0.9. At the other end where $|r|^2\sim0$, we clearly find coherence properties close to the ones of the input state $\ket{\alpha}$. A significant feature in the curve is the presence of a peak centred around $|r|^2=0.6$, which attains a measured value of $g^{(2)}(0)=2.5\pm 0.23$, above the limit of thermal states, which has been termed ``super-bunching''~\cite{Isk12}. While this is not a signature of nonclassicality {\it per se} (in principle any value can be achieved by a mixture of coherent states), we can still make inferences of nonclassical features by observing the corresponding statistics measured in Fig.~\ref{fig2:exp}. In the ideal case, the Fock state filter should significantly suppress the single photon component for a reflectivity $|r|^2=|t|^2=1/2$, by Hong-Ou-Mandel interference~\cite{San06}. Its action on a coherent state should then bring it closer to a weakly squeezed state -- which only contains contributions from even photon numbers, and can reach $g^{(2)}(0){>}3$. While our detection scheme is not able to show squeezing explicitly, these photon counting measurements shows evidence of nonclassical statistics.

\textit{Conclusions --} We have introduced a general scheme for producing a broad class of useful multi-photon states without strong optical nonlinarities, whose generation is feasible with current technology. We have exhibited how a broad range of sophisticated quantum states can be designed starting from just single photons, coherent states, beam splitters and photon counters. In conjunction with quantum memories that can drastically increase multi-photon rates, and also act as tunable beam splitters~\cite{Reim12}, photon catalysis can be used as a building block in the construction of extended quantum networks~\cite{Dat12}.

\textit{Acknowledgements --} The authors are grateful to A. Katcher for developing the detector electronics, and L. Zhang and J. Nunn for interesting discussions. This work was supported by the Engineering and Physical Sciences Research Council of the UK (EPSRC, project EP/H03031X/1), the US European Office of Aerospace Research \& Development (EOARD, project 093020), the European Commission (under Integrated Project Quantum Interfaces, Sensors, and Communication based on Entanglement (QESSENCE)) and the Royal Society. XMJ acknowledges support from NSFC(No. 11004183) and CPSF(No.201003327). MB is supported by a FASTQUAST ITN Marie Curie fellowship.


\end{document}